\documentclass[letterpaper,11pt,onecolumn]{IEEEtran}
\IEEEoverridecommandlockouts

\usepackage{graphicx}
\usepackage{psfrag}
\usepackage{subfigure}

\usepackage[cmex10]{amsmath}
\usepackage{amssymb}
\usepackage{amsthm}
\usepackage{amsfonts}

\usepackage{cases}
\usepackage{units}
\usepackage{dsfont}

\usepackage{cite}
\usepackage{url}

\newcommand{\E}{\mathrm{E}}

\DeclareMathOperator{\tr}{\mathbf{tr}}
\DeclareMathOperator{\dB}{dB}

\DeclareMathOperator{\CS}{CS}
\DeclareMathOperator{\DF}{DF}
\DeclareMathOperator{\hCS}{hCS}
\DeclareMathOperator{\hDF}{hDF}
\DeclareMathOperator{\thop}{2hop}
\DeclareMathOperator{\CF}{CF}
\DeclareMathOperator{\RD}{RD}
\DeclareMathOperator{\WZ}{WZ}

\begin{document}

\title{Transmit Signal and Bandwidth Optimization in Multiple-Antenna Relay Channels}
\author{%
	\IEEEauthorblockN{Chris~T.~K.~Ng and Gerard~J.~Foschini}\\
	\IEEEauthorblockA{Bell Laboratories, Alcatel-Lucent, Holmdel, NJ 07733\\
		Email: \{Chris.Ng, Gerard.Foschini\}@alcatel-lucent.com}
}

\maketitle
\thispagestyle{empty}

%%% ============================================================
\begin{abstract}

Transmit signal and bandwidth optimization is considered in multiple-antenna relay channels. Assuming all terminals have channel state information, the cut-set capacity upper bound and decode-and-forward rate under full-duplex relaying are evaluated by formulating them as convex optimization problems. For half-duplex relays, bandwidth allocation and transmit signals are optimized jointly. Moreover, achievable rates based on the compress-and-forward transmission strategy are presented using rate--distortion and Wyner--Ziv compression schemes. It is observed that when the relay is close to the source, decode-and-forward is almost optimal, whereas compress-and-forward achieves good performance when the relay is close to the destination.

\end{abstract}

\begin{IEEEkeywords}
Bandwidth allocation, convex optimization, relay channels, multiple-antenna, transmit covariance matrices.
\end{IEEEkeywords}

%%% ============================================================
\section{Introduction}
\label{sec:intro}

In wireless communications, the transmission rate is fundamentally limited by the channel propagation loss over the range of transmission.
Relaying has been proposed as a transmission strategy that can improve the performance of wireless systems.
In a relay channel, in addition to the source and the destination, there is also a relay terminal.
The relay does not have its own data to send or receive; its intention is to facilitate the transmission between the source and destination.
On one hand, the relay brings additional power to the network, as typically, the relay is under its own power source.
On the other hand,
a relay can also help to shorten the transmission range by enabling communication over two hops,
and cooperate with the source to perform joint encoding of the transmit signals.
However, if the relay transmission scheme is not designed properly, the relay may also create undesirable interference to the terminal at the destination.
In this paper, we investigate the optimization of transmit signals in relay channels.
In particular, we consider a multiple-input multiple-output (MIMO) relay channel, where the source, relay, and destination terminal all have multiple antennas.
In wireless communications, using multiple transmit and receive antennas has been shown to
provide substantial improvement in channel capacity \cite{foschini98:wcomm_mult_ant, telatar99:cap_mimo_gaus}.
We also evaluate the MIMO relay channel rates under different network geometry, and investigate the effectiveness of the corresponding relaying schemes.

% Related Works:
The three-node relay channel model is proposed in \cite{vdmeulen71:3t_comm_ch}.
In \cite{cover79:cap_relay}, a capacity upper bound and achievable coding strategies are presented for the relay channel,
but the relay channel capacity remains an open problem.
For Gaussian single-antenna relay channels, capacity bounds and power allocation are studied in \cite{host-madsen05:power_relay}.
Capacity bounds on half-duplex relaying are presented in \cite{khojastepour03:cap_cheap_relay, khojastepour03:gaus_cheap_relay_ch}.
Bandwidth and power allocation are considered in \cite{liang05:ortho_relay_alloc_cap, liang07:res_alloc_relay_min_max} for fading orthogonal relay channels,
and in \cite{maric04:bw_pow_relay_net} for the amplify-and-forward scheme in Gaussian relay networks.
Transmitter cooperation versus receiver cooperation in relay channels is compared in \cite{ng08:csi_pow_relay}.
Relay channel coding strategies, with extensions to relay channels with multiple terminals, are given in \cite{kramer05:coop_cap_relay}.
For relay channels with multiple-antenna terminals, bounds to the cut-set capacity upper bound and decode-and-forward rate are considered in \cite{wang05:cap_mimo_relay, ng06:snr_mimo_rates}.
In \cite{yuksel07:dmt_hdpx_relay, yuksel06:dmt_m_ant_relay}, the diversity-multiplexing tradeoff is characterized for full-duplex and half-duplex MIMO relay channels.

In this paper, we consider a multiple-antenna relay channel where the terminals have knowledge of the channel state information (CSI).
We consider optimization of the source and relay transmit signals to evaluate the MIMO cut-set capacity upper bound and the decode-and-forward achievable rate,
by formulating them as convex optimization problems.
In the case of half-duplex relaying, the bandwidth allocation and the multiple-antenna transmit signals are optimized jointly.
We also present achievable rates in MIMO relay channels using the compress-and-forward approach, under which the rate--distortion and Wyner--Ziv compression schemes are considered.

The remainder of this paper is organized as follows.
Section~\ref{sec:sysmod} presents the multiple-antenna relay channel model, and the capacity upper and lower bounds.
The cut-set bound and decode-and-forward rate optimization formulations are described in Section~\ref{sec:full_d_relay} under full-duplex assumptions,
while Section~\ref{sec:half_d_relay} considers half-duplex relaying where the relay cannot simultaneously transmit and receive in the same frequency band.
The compress-and-forward transmission strategy is studied in Section~\ref{eq:CF_relaying},
and Section~\ref{sec:conclu} concludes the paper.

\paragraph*{Notation}
In this paper, $\mathds{R}$ ($\mathds{R}_+$, $\mathds{R}_{++}$) is the set of real numbers (nonnegative, positive real numbers),
$\mathds{C}$ is the complex field, $\mathds{1}$ denotes the two-element set $\{0,1\}$.
Dimensions of vectors/matrices are indicated by superscripts.
$\mathds{H}_+^N$ is the set of $N \times N$ positive semidefinite Hermitian matrices.
$X \succ Y$ ($X \succeq Y$) means the matrix $X-Y$ is positive (semi)definite.
$I_N$ is the $N \times N$ identity matrix.
$A^T$ and $A^H$ are the transpose and conjugate transpose, respectively, of a matrix $A$.
The operators $\E[\,\cdot\,]$, $\det$, $\tr$ denote, respectively, expectation, determinant and trace.
For random variables, $x\sim\mathcal{CN}(\mu,Q)$, where $x,\mu\in \mathds{C}^N$, $Q\in\mathds{H}_+^N$,
means that $x$ is a circularly symmetric complex Gaussian random $N$-vector about mean $\mu$ with covariance matrix $Q$.

%%% ============================================================
\section{System Model}
\label{sec:sysmod}

\subsection{Channel Model}

\begin{figure}
  \centering
  \psfrag{x1}[][][0.9]{$x_1$}
  \psfrag{x2}[][][0.9]{$x_2$}
  \psfrag{y1}[][][0.9]{$y_1$}
  \psfrag{y2}[][][0.9]{$y_2$}
  \psfrag{P1}[][][0.9]{$P_1$}
  \psfrag{P2}[][][0.9]{$P_2$}
  \psfrag{M1}[][][0.9]{$M_1$}
  \psfrag{N1}[][][0.9]{$N_1$}
  \psfrag{N2,M2}[][][0.9]{$N_2,M_2$}
  \psfrag{H11}[][][0.9]{$H_{11}$}
  \psfrag{H21}[][][0.9]{$H_{21}$}
  \psfrag{H12}[][][0.9]{$H_{12}$}
  \includegraphics{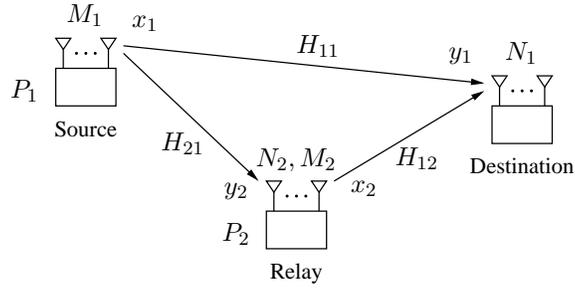}
  \caption{Multiple-antenna relay channel.}
  \label{fig:mimo_relay}
\end{figure}

Consider a three-node wireless relay channel as illustrated in Fig.~\ref{fig:mimo_relay}.
The source node wishes to send a message to the destination;
the relay node does not have its own message to send, but facilitates the transmission between the source and destination.
Suppose the source has $M_1$ transmit antennas, and the destination has $N_1$ receive antennas.
We assume the relay has $M_2$ transmit antennas and $N_2$ receive antennas
(for instance, in full-duplex operation the relay may have different sets of transmit and receive antennas).
We consider a discrete-time flat-fading channel model, which is described by
\begin{align}
\label{eq:y_1}
y_1 &= H_{11} x_1 + H_{12} x_2 + z_1\\
\label{eq:y_2}
y_2 &= H_{21} x_1 + z_2
\end{align}
where
$x_1 \in \mathds{C}^{M_1}$, $x_2 \in \mathds{C}^{M_2}$ are the respective transmit signals of the source and relay;
$y_1 \in \mathds{C}^{N_1}$, $y_2 \in \mathds{C}^{N_2}$ are the respective receive signals of the destination and relay;
and
$z_1 \sim \mathcal{CN}(0,I_{N_1}) \in \mathds{C}^{N_1}$, $z_2 \sim \mathcal{CN}(0,I_{N_2}) \in \mathds{C}^{N_2}$ are independent zero-mean circularly symmetric complex Gaussian (ZMCSCG) noise
at the destination and relay, respectively.
The complex baseband channel from the source to destination is
$H_{11} \in \mathds{C}^{N_1\times M_1}$;
from the source to relay is
$H_{21} \in \mathds{C}^{N_2\times M_1}$;
and from the relay to destination is
$H_{12} \in \mathds{C}^{N_1\times M_2}$.

We consider a block-fading channel model: the channels realize independently according to their distribution at the beginning of each fading block, and they remain unchanged within the duration of the fading block.
In this paper, we assume the channel states can be estimated accurately and conveyed timely to all terminals:
i.e., we assume channel state information (CSI) is available at all nodes.
The source and the relay are under the respective transmit power constraints: $\E[x_i^Hx_i] \leq P_i$, $i=1,2$,
where the expectations are over repeated channel uses within every fading block.
Power allocation across fading blocks is not permitted.
The transmit signals have zero mean: $\E[x_1]=0$, $\E[x_2]=0$.

It is convenient to write (\ref{eq:y_1}), (\ref{eq:y_2}) in block-matrix form
\begin{align}
y &= H x + z
\end{align}
where
\begin{align}
y &\triangleq \begin{bmatrix}y_1\\y_2\end{bmatrix} \in \mathds{C}^N, &
H &\triangleq \begin{bmatrix}H_{11} & H_{12}\\ H_{21} & 0\end{bmatrix} \in \mathds{C}^{N\times M}, &
x &\triangleq \begin{bmatrix}x_1\\x_2\end{bmatrix} \in \mathds{C}^M, &
z &\triangleq \begin{bmatrix}z_1\\z_2\end{bmatrix} \in \mathds{C}^N
\end{align}
with $M \triangleq M_1 + M_2$, and $N \triangleq N_1 + N_2$.
Moreover, we denote the joint covariance matrix of the transmit signals of the source and relay as
\begin{align}
\label{eq:Q11Q22_Q}
\begin{bmatrix}Q_{11} & Q_{12} \\ Q_{21} & Q_{22}\end{bmatrix} \triangleq Q &\triangleq \E[xx^H] \in \mathds{H}^{M}_+
\end{align}
where the conformally partitioned blocks (with respect to $x_1,x_2$) of $Q$ have dimensions
\begin{align}
Q_{11} \triangleq \E[x_1x_1^H] &\in \mathds{H}_+^{M_1},&
Q_{22} \triangleq \E[x_2x_2^H] &\in \mathds{H}_+^{M_2},&
Q_{12} \triangleq \E[x_1x_2^H] = Q_{21}^H &\in \mathds{C}^{M_1\times M_2}.
\end{align}

\subsection{Capacity Bounds and Achievable Rates}

The capacity of a relay channel is, in general, an open problem; however, there are known upper and lower bounds.
The cut-set bound described in \cite{cover79:cap_relay, cover91:eoit} provides an upper bound to the relay channel
capacity.
Intuitively, the cut-set bound states that, over any possible joint source-relay transmit signals, the relay channel capacity cannot exceed the smaller of
i) the maximum rate at which information can flow out of the source,
and ii) the maximum rate at which information can flow into the destination.

On the other hand, capacity lower bounds of the relay channel, as achieved by two coding strategies, are given in \cite{cover79:cap_relay}.
In the decode-and-forward strategy \cite[Thm.~1]{cover79:cap_relay}, transmission is done in blocks:
the relay first fully decodes the message from the source in one block,
then in the ensuing block, the relay and the source cooperatively transmit the message to the destination.
In the compress-and-forward strategy \cite[Thm.~6]{cover79:cap_relay}, the relay does not decode the source's message but sends a compressed version
of its observed signal to the destination.
The destination then combines the compressed signal with its own receive signal to decode the source's message.
A detailed discussion on the relay channel coding strategies can be found in \cite{kramer05:coop_cap_relay}.

In the following sections, we evaluate these capacity bounds and achievable rates for full- and half-duplex Gaussian MIMO relay channels.
Under half-duplex relaying in Section~\ref{sec:half_d_relay}, we impose the constraint that the relay cannot simultaneously transmit and receive in the same frequency band.
For performance comparisons, we also consider: direct transmission when the relay is not available
(i.e., the capacity of an $M_1\times N_1$ MIMO channel under transmit power constraint $P_1$);
orthogonal two-hop relaying (Section~\ref{sec:two_hop_relay}); and the scenario when the relay is co-located with the source or destination (Section~\ref{sec:relay_mimo_cap}).

%%% ============================================================
\section{MIMO Relay Channel Capacity Bounds}
\label{sec:full_d_relay}

In this section, we present the optimization frameworks for evaluating the MIMO relay channel cut-set capacity upper bound and the decode-and-forward achievable rate.
We first adopt the full-duplex assumption where the relay can transmit and receive in the same frequency band at the same time.
In practice, full-duplex transmission is difficult to realize.
Nevertheless, the full-duplex model provides insight into the design of effective coding strategies for relay channels,
and serves as a performance upper bound for half-duplex systems.
Half-duplex relaying is considered in Section~\ref{sec:half_d_relay} where the relay cannot simultaneously transmit and receive in the same band.

\subsection{Cut-Set Capacity Upper Bound}
\label{sec:full_cutset}

The cut-set capacity upper bound \cite{cover79:cap_relay} for the relay channel is given as an optimization in terms of the channel mutual information as follows:
\begin{align}
\label{eq:R_CS_x1_x2}
R_{\CS} &= \max_{p(x_1,x_2)} \;\min\bigl\{I(x_1;y_1,y_2|x_2),\, I(x_1,x_2;y_1)\bigr\}\\
\label{eq:R_CS_x}
&= \max_{p(x)} \;\min\bigl\{I(x_1;y|x_2),\, I(x;y_1)\bigr\}.
\end{align}
Gaussian signals are optimal in the cut-set bound and decode-and-forward rate \cite[Proposition~2]{kramer05:coop_cap_relay}.
We denote the transmit signals by:
$x\sim\mathcal{CN}(0,Q)$, where $Q \in \mathds{H}_+^M$ is the covariance matrix of $x$.
Then the mutual information expressions in (\ref{eq:R_CS_x}) evaluate \cite{telatar99:cap_mimo_gaus} to
\begin{align}
\label{eq:R_CS_logdet_Q}
R_{\CS} &= \max_{Q\;:\;\tr Q_{ii} \leq P_i,\; i=1,2} \;\min\bigl\{\log\det(I_N + H_1 Q_{1|2} H_1^H),\, \log\det(I_{N_1} + \tilde{H}_1 Q \tilde{H}_1^H)\bigr\}
\end{align}
where
$H_1$ and $\tilde{H}_1$ are, respectively, the first block column and block row of $H$
\begin{align}
H_1 &\triangleq \begin{bmatrix}H_{11}\\ H_{21}\end{bmatrix} \in \mathds{C}^{N\times M_1}, &
\tilde{H}_1 &\triangleq \begin{bmatrix}H_{11} & H_{12}\end{bmatrix} \in \mathds{C}^{N_1\times M}
\end{align}
and the conditional covariance matrix $Q_{1\!\vert2} \triangleq \E[x_1x_1^H \vert x_2]$ is given by the Schur complement of $Q_{22}$ in $Q$
\begin{align}
\label{eq:Q12_Schur}
Q_{1|2} = Q_{11} - Q_{12}Q_{22}^{-1}Q_{21}
\end{align}
where we assume $Q_{22}\succ0$.
The zero-mean Gaussian signal $x$ is fully characterized by its covariance; therefore, in (\ref{eq:R_CS_logdet_Q}), the sole optimization variable is the joint covariance matrix $Q$.

The cut-set bound maximization in (\ref{eq:R_CS_logdet_Q}) can be formulated as the following optimization problem:
\begin{align}
\label{eq:CS_max_RCS}
\text{maximize}\quad & R_{\CS}\\
\label{eq:CS_over_RQ}
\text{over}\quad & R_{\CS}\in\mathds{R}_+,\; Q\in\mathds{H}_+^{M},\; Q_{1|2}\in\mathds{H}_+^{M_1}\\
\label{eq:CS_R_logdet_Q12}
\text{subject to}\quad
& R_{\CS} \leq \log\det(I_N + H_1 Q_{1|2} H_1^H)\\
\label{eq:CS_R_logdet_Q}
& R_{\CS} \leq \log\det(I_{N_1} + \tilde{H}_1 Q \tilde{H}_1^H)\\
\label{eq:CS_C1QC1_P1}
& \tr(C_1^T Q C_1) \leq P_1\\
\label{eq:CS_C2QC2_P2}
& \tr(C_2^T Q C_2) \leq P_2\\
\label{eq:CS_Q_CQC}
& Q - C_1 Q_{1|2} C_1^T \succeq 0
\end{align}
where $C_1,C_2$ are constant matrices defined as
\begin{align}
\label{eq:C1_C2_def}
C_1 &\triangleq \begin{bmatrix}I_{M_1}\\ 0\end{bmatrix} \in \mathds{1}^{M\times M_1},&
C_2 &\triangleq \begin{bmatrix}0\\I_{M_2}\end{bmatrix} \in \mathds{1}^{M\times M_2}.
\end{align}
In the optimization, (\ref{eq:CS_R_logdet_Q12}), (\ref{eq:CS_R_logdet_Q}) follow from the two terms inside the $\min$ expression in (\ref{eq:R_CS_logdet_Q});
and (\ref{eq:CS_C1QC1_P1}), (\ref{eq:CS_C2QC2_P2}) represent the per-node transmit power constraints at the source and relay, respectively.
The constraint (\ref{eq:CS_Q_CQC}) results from relaxing the equality constraint in (\ref{eq:Q12_Schur})
\begin{align}
\label{eq:Q_eq_ineq}
Q_{1|2} = Q_{11} - Q_{12}Q_{22}^{-1}Q_{21} \quad\Longrightarrow\quad
Q_{1|2} \preceq Q_{11} - Q_{12}Q_{22}^{-1}Q_{21}.
\end{align}
By the semidefiniteness property of Schur complements \cite{zhang05:schur_cmpm_app},
we have the following identity on the right-hand side of (\ref{eq:Q_eq_ineq}):
\begin{align}
\label{eq:Q_Schur_eq_block}
(Q_{11}-Q_{1|2}) - Q_{12}Q_{22}^{-1}Q_{21} \succeq 0
\quad\Longleftrightarrow\quad
\begin{bmatrix}
(Q_{11}-Q_{1|2}) & Q_{12}\\
Q_{21} & Q_{22}
\end{bmatrix}\succeq 0
\end{align}
where the right-hand side of (\ref{eq:Q_Schur_eq_block})
is equivalent to
$Q - C_1 Q_{1|2} C_1^T \succeq 0$ when written in the block-matrix form as defined in (\ref{eq:Q11Q22_Q}).
Finally, we show the relaxation in (\ref{eq:Q_eq_ineq}) does not increase the optimal value in (\ref{eq:CS_max_RCS}).
Suppose given a set of fixed $Q_{11},Q_{12},Q_{21},Q_{22}$, we consider all $X\in\mathds{H}_+^{M_1}$ such that
$X\preceq Q_{11} - Q_{12}Q_{22}^{-1}Q_{21}$.
Recalling that the determinant is matrix increasing \cite{boyd04:convex_opt} on the set of positive semidefinite matrices, we get
\begin{align}
\log\det(I_N + H_1 X H_1^H) \leq \log\det\bigl(I_N + H_1 (Q_{11} - Q_{12}Q_{22}^{-1}Q_{21}) H_1^H\bigr)
\end{align}
which only limits the feasible set in (\ref{eq:CS_R_logdet_Q12}).

The maximization in (\ref{eq:CS_max_RCS})--(\ref{eq:CS_Q_CQC}) is a convex optimization problem;
in particular, the log-determinant function is concave on positive definite matrices \cite{boyd04:convex_opt}.
The solution of (\ref{eq:CS_max_RCS}) can be efficiently computed using standard convex optimization numerical techniques,
for instance, by the interior-point method \cite{boyd04:convex_opt, renegar01:math_ipm_cvxopt}.
The above optimization can also be solved by the CVX \cite{grant08:graph_nonsm_cvx, grant09:cvx_dcp_web} software package,
which uses a successive approximation approach to model the log-determinant inequalities.
All optimization formulations presented in this paper are convex problems, unless otherwise noted.

In addition to the per-node power constraints (\ref{eq:CS_C1QC1_P1}), (\ref{eq:CS_C2QC2_P2}),
if the source and relay are also under per-antenna power constraints, they can be readily incorporated in the convex optimization formulation.
Let the antenna power constraints of the source and relay, respectively, be $p_{1,1},\dotsc,p_{1,M_1}$ and $p_{2,1},\dotsc,p_{1,M_2}$.
The per-antenna power constraints are represented by
\begin{align}
q_{i,i} &\leq p_{1,i},\quad i = 1,\dots,M_1,&
q_{M_1+j,M_1+j} &\leq p_{2,j},\quad j = 1,\dots,M_2
\end{align}
where $q_{i,i}$ is the $(i,i)$ entry of the covariance matrix $Q$, with $i=1,\dotsc,M$.

\subsection{Decode-and-Forward Achievable Rate}
\label{sec:full_decfor}

The decode-and-forward \cite[Thm.~1]{cover79:cap_relay} relay channel achievable rate is given by
\begin{align}
R_{\DF} &= \max_{p(x_1,x_2)} \;\min\bigl\{I(x_1;y_2|x_2),\, I(x_1,x_2;y_1)\bigr\}\\
&= \max_{p(x)} \;\min\bigl\{I(x_1;y_2|x_2),\, I(x;y_1)\bigr\}\\
&= \max_{Q\;:\;\tr Q_{ii} \leq P_i,\; i=1,2} \;\min\bigl\{\log\det(I_{N_2} + H_{21} Q_{1|2} H_{21}^H),\, \log\det(I_{N_1} + \tilde{H}_1 Q \tilde{H}_1^H)\bigr\}
\end{align}
where $Q_{1|2}$ is as given in (\ref{eq:Q12_Schur}).
The decode-and-forward rate can be formulated as the solution to the following convex optimization problem:
\begin{align}
\label{eq:DF_max_R_DF}
\text{maximize}\quad & R_{\DF}\\
\text{over}\quad & R_{\DF}\in\mathds{R}_+,\; Q\in\mathds{H}_+^{M},\; Q_{1|2}\in\mathds{H}_+^{M_1}\\
\label{eq:DF_RDF_logdet_Q12}
\text{subject to}\quad
& R_{\DF} \leq \log\det(I_{N_2} + H_{21} Q_{1|2} H_{21}^H)\\
& R_{\DF} \leq \log\det(I_{N_1} + \tilde{H}_1 Q \tilde{H}_1^H)\\
& \tr(C_1^T Q C_1) \leq P_1\\
& \tr(C_2^T Q C_2) \leq P_2\\
\label{eq:DF_Q_CQ12C}
& Q - C_1 Q_{1|2} C_1^T \succeq 0.
\end{align}
The derivations for the optimization problem formulation parallel those presented in the previous section.
Note that the decode-and-forward optimization problem (\ref{eq:DF_max_R_DF})--(\ref{eq:DF_Q_CQ12C}) is similar to the
cut-set bound optimization problem (\ref{eq:CS_max_RCS})--(\ref{eq:CS_Q_CQC}).
The only difference is the aggregate channel $H_1$ in (\ref{eq:CS_R_logdet_Q12}) versus the source-relay channel $H_{21}$ in (\ref{eq:DF_RDF_logdet_Q12}).
Consequently, we expect the decode-and-forward scheme is nearly capacity-achieving when the direct channel $H_{11}$ is weak relative to the source-relay channel $H_{21}$.

\subsection{Numerical Results}
\label{sec:full_num_res}

\begin{figure}
  \centering
  \psfrag{dx}[][][0.9]{$d_x$}
  \psfrag{dy}[][][0.9]{$d_y$}
  \psfrag{(0,0)}[][][0.9]{$(0,0)$}
  \psfrag{(1,0)}[][][0.9]{$(1,0)$}
  \includegraphics{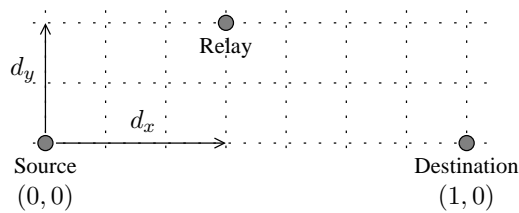}
  \caption{Geometry of the source, relay, and destination nodes.}
  \label{fig:relay_geometry}
\end{figure}

In the numerical examples in this paper, we assume a network geometry as depicted in Fig.~\ref{fig:relay_geometry}.
In the two-dimensional network, the source is located at coordinates $(0,0)$, the destination is at $(1,0)$, and the relay is at $(d_x,d_y)$.
We will use a distance-based path-loss power attenuation exponent $\eta=4$, combined with independent and identically distributed (i.i.d.) Rayleigh fading for each channel matrix entry.
Shadow fading can also be included, but it is omitted here to allow a simple geometric interpretation of the relay network topology.
The relay channel matrices are given as
\begin{align}
\label{eq:H_Hw_dx_dy}
H_{11} &= H_w^{(1)},&
H_{21} &= (d_x^2+d_y^2)^{-\eta/4} H_w^{(2)},&
H_{12} &= \bigl((1-d_x)^2+d_y^2\bigr)^{-\eta/4} H_w^{(3)}
\end{align}
where $H_w^{(1)} \in \mathds{C}^{N_1 \times M_1}$, $H_w^{(2)} \in \mathds{C}^{N_2 \times M_1}$, $H_w^{(3)} \in \mathds{C}^{N_1 \times M_3}$,
with each entry of $H_w^{(1)}$, $H_w^{(2)}$, $H_w^{(3)}$ i.i.d. $\sim\mathcal{CN}(0,1)$.
For the numerical experiments presented in this paper, 50 random instances of the channel realizations $H_w^{(1)}$, $H_w^{(2)}$, $H_w^{(3)}$ are generated.
Then under each channel realization, the corresponding optimization problems are solved to evaluate the relay channel capacity bounds and achievable rates.
For consistent comparison, the same set of channel realizations is used to compute the performance of the different coding schemes under consideration.
The convex optimization problems are solved using the barrier interior-point algorithm described in \cite[Section~11.3]{boyd04:convex_opt}.

\begin{figure}
  \centering
  \includegraphics*[width=10.85cm]{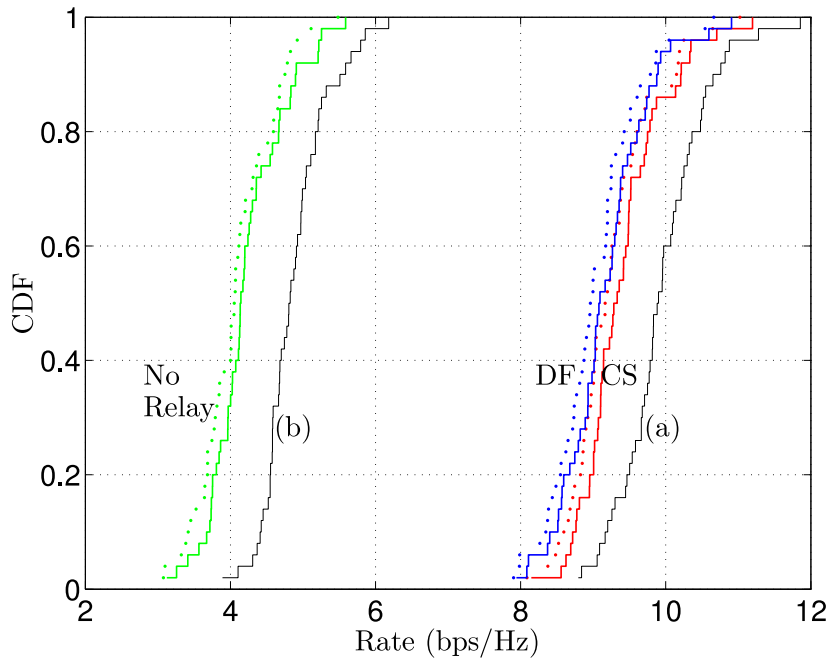}
  \caption{Full-duplex CDFs of MIMO relay channel cut-set (CS) bound and decode-and-forward (DF) rate.
  The relay is located at $(\nicefrac{1}{3},\nicefrac{1}{2})$.
  All terminals have four antennas: $M_1=N_1=M_2=N_2=4$, and unit power constraints: $P_1=P_2=0\,\dB$.
  The dotted lines represent the rates under per-antenna power constraints.
  The plots (a) and (b) correspond to the upper and lower bounds, respectively, from \cite[Thms.~3.1 and 3.2]{wang05:cap_mimo_relay}.}
  \label{fig:relay_Mi4_R50_dx03_dy05}
\end{figure}

The empirical cumulative distribution functions (CDFs) of the cut-set (CS) bound and decode-and-forward (DF) rate of a MIMO relay channel,
where the relay is located at $(d_x,d_y) = (\nicefrac{1}{3},\nicefrac{1}{2})$, are shown in Fig.~\ref{fig:relay_Mi4_R50_dx03_dy05} (solid lines).
All terminals have four antennas: $M_1=N_1=M_2=N_2=4$, and unit power constraints: $P_1=P_2=0\,\dB$.
The dotted lines represent the rates under per-antenna power constraints, where all antennas have equal constraints:
$p_{1,1} = \dotsb = p_{1,M_1} = P_1/M_1$, and $p_{2,1} = \dotsb = p_{2,M_2} = P_2/M_2$.
Also shown in the plot is the MIMO capacity of the direct channel $H_{11}$ when the relay node is not available (No Relay).
The decode-and-forward achievable rate considerably outperforms the direct channel capacity and is quite close to the cut-set capacity upper bound.
For all coding schemes, imposing per-antenna power constraints only slightly reduces the rates as compared to per-node power constraints.
For comparison, the upper and lower bounds from \cite[Thms.~3.1 and 3.2]{wang05:cap_mimo_relay} are plotted and labeled (a) and (b), respectively.
It is observed that the capacity upper and lower bounds can be tightened when the transmit signals of the source and relay are optimized.
The upper bound (a) is computed by searching over combinations of
$\rho=\{0,0.05,\dotsc,0.9,0.95\}$ and $a=\{10^{-1},10^{-0.9},\dotsc,10^{0.9},10^1\}$, where the parameters $\rho,a$ are as defined in \cite{wang05:cap_mimo_relay}.
For each choice of $\rho,a$, a convex optimization problem is solved.

%%% ============================================================
\section{Half-Duplex Relaying}
\label{sec:half_d_relay}

In Section~\ref{sec:full_d_relay}, we assumed the relay was able to transmit and receive simultaneously in the same band.
Such full-duplex radios can be difficult to implement in practice.
In this section, we consider a half-duplex relay, where the relay receives in one band and transmits over a different band.
In particular, we assume the channel has unit bandwidth, and it is partitioned into sub-channel Band~1 with bandwidth~$w_1$,
and another orthogonal sub-channel Band~2 with bandwidth~$w_2$, with $w_1+w_2 \leq 1$.
The relay can only receive in Band~1 and it can only transmit in Band~2.
Hence the channel is described by
\begin{align}
y_1^{(1)} &= H_{11} x_1^{(1)} + z_1^{(1)},&
y_1^{(2)} &= H_{11} x_1^{(2)} + H_{12}x_2^{(2)} + z_1^{(2)}\\
y_2^{(1)} &= H_{21} x_1^{(1)} + z_2^{(2)},&
y_2^{(2)} &= 0
\end{align}
where the superscripts designate the corresponding bands.
The noise powers in the sub-channels are given by
\begin{align}
\E[z_i^{(1)}(z_i^{(1)})^H] &= w_1 I_{N_i},&
\E[z_i^{(2)}(z_i^{(2)})^H] &= w_2 I_{N_i},&
i&=1,2.
\end{align}
Let $Q_{11}^{(1)}$, $Q^{(2)}$ be the transmit signal covariance matrices in the two bands
\begin{align}
Q_{11}^{(1)} &\triangleq \E[x_1^{(1)}(x_1^{(1)})^H] \in\mathds{H}_+^{M_1},&
Q^{(2)} &\triangleq \E[x^{(2)}(x^{(2)})^H] \in\mathds{H}_+^M,&
x^{(2)} &\triangleq [x_1^{(2)} \; x_2^{(2)}]^T \in\mathds{C}^M.
\end{align}
We assume the bandwidth allocation and the transmit signal covariances in each band can be optimized with respect to the channel realizations.
In the following sections, we consider the cut-set capacity upper bound and the achievable rates under the half-duplex relaying constraint.

\subsection{Half-Duplex Cut-Set Bound}
\label{sec:half_cutset}

\begin{figure}
  \centering
  \subfigure[Cut set around the source.]{
  \psfrag{w1}[][][0.9]{$w_1$}
  \psfrag{w2}[][][0.9]{$w_2$}
  \psfrag{R1}[][][0.9]{$R_1$}
  \psfrag{R2}[][][0.9]{$R_2$}
  \includegraphics{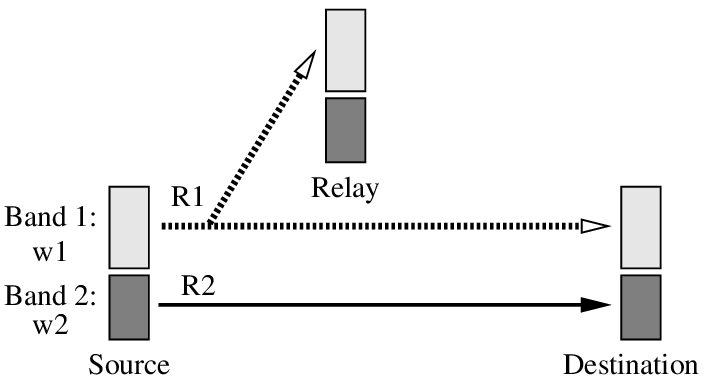}
  \label{fig:half_relay_CS_src}
  }
  \subfigure[Cut set around the destination.]{
  \psfrag{w1}[][][0.9]{$w_1$}
  \psfrag{w2}[][][0.9]{$w_2$}
  \psfrag{Rc}[][][0.9]{$R_{\mathrm{c}}$}
  \psfrag{Rd}[][][0.9]{$R_{\mathrm{d}}$}
  \includegraphics{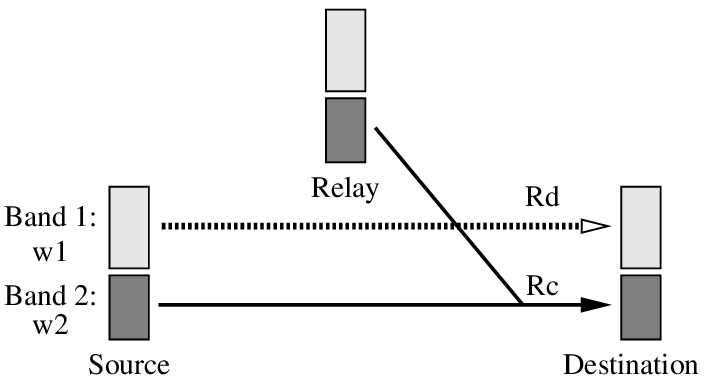}
  \label{fig:half_relay_CS_dst}
  }
  \caption{Half-duplex cut-set bound.}
  \label{fig:half_relay_CS}
\end{figure}

Let the mutual information across the different cut sets be designated as labeled in Fig.~\ref{fig:half_relay_CS}.
The cut set around the source is shown in Fig.~\ref{fig:half_relay_CS_src}.
Let $R_1,R_2$, respectively, denote the egress information rate out of the source in Band~1 and Band~2.
On the other hand, for the cut set around the destination shown in Fig.~\ref{fig:half_relay_CS_dst},
let $R_{\mathrm{d}},R_{\mathrm{c}}$ be the ingress information rate into the destination in Band~1 and Band~2, respectively.
Optimizing over the transmit signals and the bandwidth allocation,
the half-duplex cut-set bound is characterized as follows:
\begin{align}
\text{maximize}\quad & R_{\hCS}\\
\text{over}\quad & R_{\hCS},R_1,R_2,R_{\mathrm{d}},R_{\mathrm{c}},w_1,w_2\in\mathds{R}_+,\; Q_{11}^{(1)}\in\mathds{H}_+^{M_1},\; Q^{(2)}\in\mathds{H}_+^M\\
\text{subject to}\quad
& R_{\hCS} \leq \min(R_1+R_2,\, R_{\mathrm{d}}+R_{\mathrm{c}})\\
\label{eq:hCS_R1}
& R_1 \leq w_1 \log\det\bigl(I_N + \tfrac{1}{w_1} H_1 Q_{11}^{(1)} H_1^H)\\
\label{eq:hCS_R2}
& R_2 \leq w_2 \log\det\bigl(I_{N_1} + \tfrac{1}{w_2} H_{11} C_1^TQ^{(2)}C_1 H_{11}^H\bigr)\\
\label{eq:hCS_Rd}
& R_{\mathrm{d}} \leq w_1 \log\det\bigl(I_{N_1} + \tfrac{1}{w_1} H_{11}Q_{11}^{(1)}H_{11}^H\bigr)\\
\label{eq:hCS_Rc}
& R_{\mathrm{c}} \leq w_2 \log\det\bigl(I_{N_1} + \tfrac{1}{w_2} \tilde{H}_1Q^{(2)}\tilde{H}_1^H)\\
& \tr Q_{11}^{(1)} + \tr (C_1^T Q^{(2)}C_1) \leq P_1\\
& \tr(C_2^T Q^{(2)}C_2) \leq P_2\\
& w_1 + w_2 \leq 1
\end{align}
where $C_1,C_2$ are as defined in (\ref{eq:C1_C2_def}).
By continuity we define: $w\log\det(I+X/w)\vert_{w=0} \triangleq 0$, for all $X\succeq0$.
The right-hand side of each constraint in (\ref{eq:hCS_R1}), (\ref{eq:hCS_R2}), (\ref{eq:hCS_Rd}), (\ref{eq:hCS_Rc})
is a concave function, being the perspective of the log-determinant function.
(Given a function $f(x)$, the perspective of $f$ is defined as the function $g(x,t)=tf(x/t)$, $t\in\mathds{R}_{++}$,
and the perspective operation preserves convexity \cite{boyd04:convex_opt}.)

\subsection{Half-Duplex Decode-and-Forward Rate}
\label{sec:half_decfor}

\begin{figure}
  \centering
  \psfrag{w1}[][][0.9]{$w_1$}
  \psfrag{w2}[][][0.9]{$w_2$}
  \psfrag{Rc}[][][0.9]{$R_{\mathrm{c}}$}
  \psfrag{Rr}[][][0.9]{$R_{\mathrm{r}}$}
  \psfrag{Rd}[][][0.9]{$R_{\mathrm{d}}$}
  \includegraphics{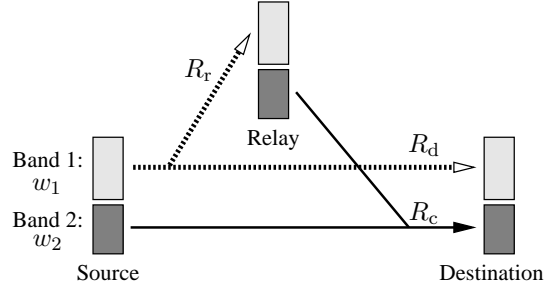}
  \caption{Half-duplex decode-and-forward.}
  \label{fig:half_relay_DF}
\end{figure}

Fig.~\ref{fig:half_relay_DF} depicts the operation of decode-and-forward in the half-duplex mode.
In Band~1, the source sends to the relay at rate $R_{\mathrm{r}}$, of which $R_{\mathrm{d}}$ is decodable at the destination.
The relay fully decodes the message from the source, and in Band~2 the source and relay cooperatively send to the destination additional information at rate $R_{\mathrm{c}}$.
The half-duplex decode-and-forward optimization is given as follows:
\begin{align}
\text{maximize}\quad & R_{\hDF}\\
\text{over}\quad & R_{\hDF},R_{\mathrm{r}},R_{\mathrm{d}},R_{\mathrm{c}},w_1,w_2\in\mathds{R}_+,\; Q_{11}^{(1)}\in\mathds{H}_+^{M_1},\; Q^{(2)}\in\mathds{H}_+^M\\
\text{subject to}\quad
& R_{\hDF} \leq \min(R_{\mathrm{r}},\,R_{\mathrm{d}}+R_{\mathrm{c}})\\
& R_{\mathrm{r}} \leq w_1 \log\det\bigl(I_{N_2} + \tfrac{1}{w_1} H_{21}Q_{11}^{(1)}H_{21}^H\bigr)\\
& R_{\mathrm{d}} \leq w_1 \log\det\bigl(I_{N_1} + \tfrac{1}{w_1} H_{11}Q_{11}^{(1)}H_{11}^H\bigr)\\
& R_{\mathrm{c}} \leq w_2 \log\det\bigl(I_{N_1} + \tfrac{1}{w_2} \tilde{H}_1Q^{(2)}\tilde{H}_1^H\bigr)\\
& \tr Q_{11}^{(1)} + \tr(C_1^TQ^{(2)}C_1) \leq P_1\\
& \tr(C_2^TQ^{(2)}C_2) \leq P_2\\
& w_1 + w_2 \leq 1.
\end{align}

\subsection{Two-Hop Relaying}
\label{sec:two_hop_relay}

\begin{figure}
  \centering
  \psfrag{w1}[][][0.9]{$w_1$}
  \psfrag{w2}[][][0.9]{$w_2$}
  \psfrag{Rsr}[][][0.9]{$R_{\mathrm{sr}}$}
  \psfrag{Rrd}[][][0.9]{$R_{\mathrm{rd}}$}
  \includegraphics{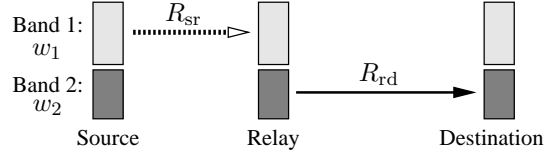}
  \caption{Two-hop relaying.}
  \label{fig:relay_2hop}
\end{figure}

Two-hop relaying is a simple scheme that imposes relatively small coordination overhead between the source and relay.
Its operation is portrayed in Fig.~\ref{fig:relay_2hop}.
In Band~1, the source transmits to the relay with signal covariance $Q_{11}^{(1)}$.
The relay then decodes the message from the source, and re-encodes it to transmit to the destination in Band~2 with covariance $Q_{22}^{(2)} \in \mathds{H}_+^{M_2}$.
The following rate is achievable
\begin{align}
\text{maximize}\quad & R_{\thop}\\
\text{over}\quad & R_{\thop},R_{\mathrm{sr}},R_{\mathrm{rd}},w_1,w_2\in\mathds{R}_+,\; Q_{11}^{(1)}\in\mathds{H}_+^{M_1},\; Q_{22}^{(2)}\in\mathds{H}_+^{M_2}\\
\text{subject to}\quad
& R_{\thop} \leq \min(R_{\mathrm{sr}},\,R_{\mathrm{rd}})\\
& R_{\mathrm{sr}} \leq w_1 \log\det\bigl(I_{N_2} + \tfrac{1}{w_1} H_{21}Q_{11}^{(1)}H_{21}^H\bigr)\\
& R_{\mathrm{rd}} \leq w_2 \log\det\bigl(I_{N_1} + \tfrac{1}{w_2} H_{12}Q_{22}^{(2)}H_{12}^H\bigr)\\
& \tr Q_{11}^{(1)} \leq P_1\\
& \tr Q_{22}^{(2)} \leq P_2\\
& w_1 + w_2 \leq 1
\end{align}
where $R_{\mathrm{sr}}$ represents the rate from the source to the relay,
and $R_{\mathrm{rd}}$ represents the rate from the relay to the destination.

\subsection{Co-Location MIMO Capacity}
\label{sec:relay_mimo_cap}

For comparison, we also consider the performance of the relay channel when the relay is close to the source or the destination.
When the relay is co-located with the source, where it can cooperate perfectly with the source without overhead,
the resulting channel is equivalent to an $M\times N_1$ MIMO channel.
In the MIMO channel, $M_1$ of the transmit antennas are under a sum power constraint of $P_1$,
and $M_2$ antennas are under sum power constraint $P_2$.
The MIMO capacity is given by
\begin{align}
\label{eq:R_M_N1}
\text{maximize}\quad & R_{M\times N_1}\\
\text{over}\quad & R_{M\times N_1}\in\mathds{R}_+,\; Q\in\mathds{H}_+^{M}\\
\text{subject to}\quad
& R_{M\times N_1} \leq \log\det(I_{N_1} + \tilde{H}_1 Q \tilde{H}_1^H)\\
& \tr(C_1^T Q C_1) \leq P_1\\
& \tr(C_2^T Q C_2) \leq P_2.
\end{align}
The above maximization is a convex optimization problem,
and can be solved by the software package SDPT3 \cite{tutuncu03:sdpt3}, which directly supports the log-determinant construct in the optimization objective function.

On the other hand, when the relay is co-located with the destination, we assume they can cooperate perfectly without overhead.
In this case, the resulting channel is equivalent to an $M_1\times N$ MIMO channel under a transmit power constraint of $P_1$.
The MIMO channel capacity is
\begin{align}
\label{eq:R_M1_N}
R_{M_1\times N} &= \max_{\tr Q_{11} \leq P_1} \;
\log\det(I_N + H_1 Q_{11} H_1^H)
\end{align}
where the solution is given by waterfilling power allocation \cite{paulraj03:intro_st_wcom} along the eigenmodes of $H_1^HH_1$.

\subsection{Numerical Results}
\label{sec:half_num_res}

\begin{figure}
  \centering
  \includegraphics*[width=10.85cm]{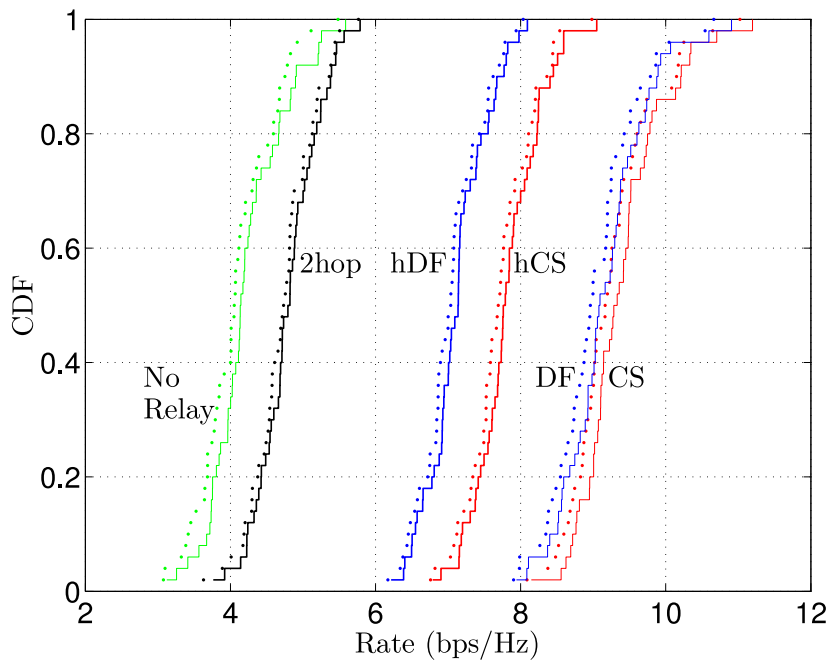}
  \caption{Half-duplex CDFs of MIMO relay channel cut-set (hCS) bound, decode-and-forward (hDF) rate, and two-hop (2hop) relaying rate.
  The relay is located at $(\nicefrac{1}{3},\nicefrac{1}{2})$.
  All terminals have four antennas: $M_1=N_1=M_2=N_2=4$, and unit power constraints: $P_1=P_2=0\,\dB$.
  The dotted lines represent the rates under per-antenna power constraints.}
  \label{fig:relay_half_Mi4_R50_dx03_dy05}
\end{figure}

Fig.~\ref{fig:relay_half_Mi4_R50_dx03_dy05} shows the empirical CDF of the half-duplex cut-set (hCS) bounds, the half-duplex decode-and-forward (hDF) rates, and the two-hop relaying (2hop) rates for the MIMO relay channel with parameters as described in Section~\ref{sec:full_num_res}.
Again, the dotted lines represent the rates under per-antenna power constraints.
For comparison, Fig.~\ref{fig:relay_half_Mi4_R50_dx03_dy05} also includes the full-duplex cut-set bound, full-duplex decode-and-forward rate,
and the MIMO capacity of the direct channel $H_{11}$ without the relay.
It is observed that the half-duplex rates fall moderately as compared to the full-duplex rates.
Moreover, the gap between the decode-and-forward rate and the cut-set capacity upper bound widens under the half-duplex mode.
On the other hand, half-duplex decode-and-forward still provides a sizable capacity gain over direct transmission,
while the two-hop relaying scheme achieves only marginally higher rates than when the relay is not available.
Similar to the full-duplex case, imposing the per-antenna power constraints reduces the rates only slightly.

\begin{figure}
  \centering
  \includegraphics*[width=10.85cm]{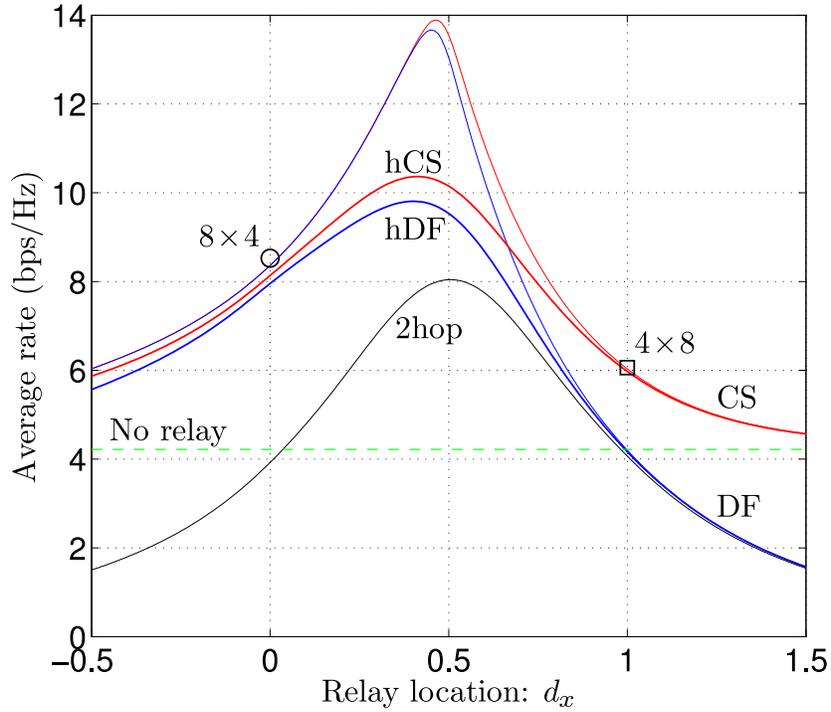}
  \caption{Full- and half-duplex cut-set (CS) bound, decode-and-forward (DF) rate, and two-hop (2hop) relaying rate with respect to the relay location.}
  \label{fig:relay_dist01_Mi4_R50_co}
\end{figure}

\begin{figure}
  \centering
  \includegraphics*[width=10.85cm]{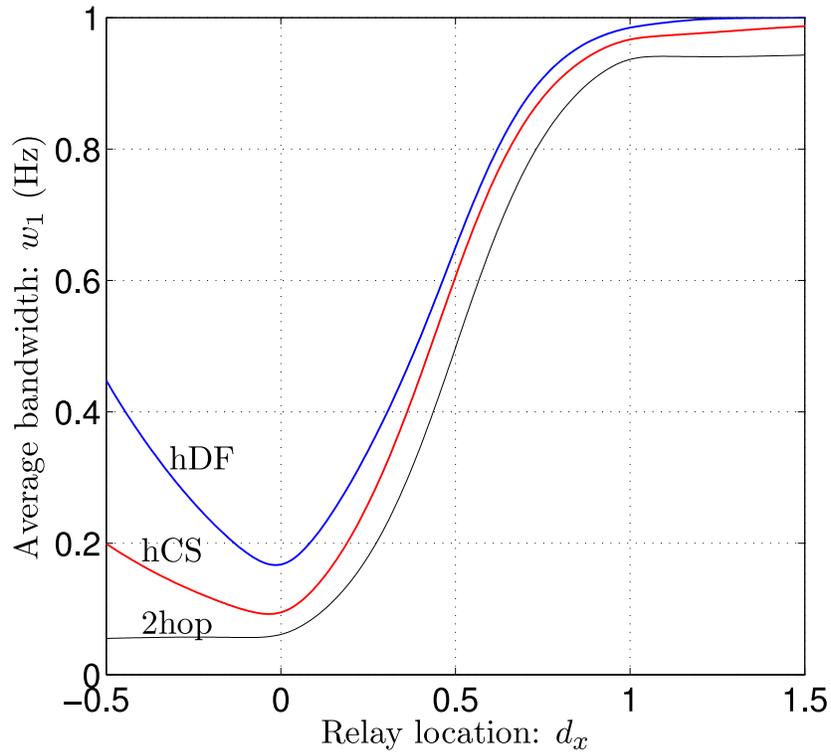}
  \caption{Bandwidth allocation in the half-duplex relaying schemes with respect to the relay location.}
  \label{fig:relay_dist01_BW_Mi4_R50}
\end{figure}

Next, we investigate the relay channel capacity bound and achievable rate as a function of the relay position.
In the following numerical experiments, we fix $d_y=\nicefrac{1}{10}$, and vary $d_x$ from $\nicefrac{-1}{2}$ to $1\nicefrac{1}{2}$;
therefore, the relay ranges from being closer to the source, to being closer to the destination.
Again, all terminals in the network have four antennas: $M_1=N_1=M_2=N_2=4$, and unit power constraints: $P_1=P_2=0\,\dB$.
The average rates for the different full- and half-duplex relaying schemes are plotted in Fig.~\ref{fig:relay_dist01_Mi4_R50_co};
they are computed over the 50 sets of random channel realizations with distance-based path-loss as given in (\ref{eq:H_Hw_dx_dy}).
The $8\times 4$ MIMO capacity given by (\ref{eq:R_M_N1}), corresponding to the case where the relay is co-located with the source, is indicated by a circle,
whereas the $4\times 8$ MIMO capacity (\ref{eq:R_M1_N}), under relay-destination co-location, is indicated by a square.
For the half-duplex schemes, the bandwidth allocation in terms of the relay location is shown in Fig.~\ref{fig:relay_dist01_BW_Mi4_R50}.

Under the full-duplex mode of operation, the decode-and-forward relaying scheme offers substantial capacity gain over transmission using only the direct channel.
Over a wide range when the relay is close to the source, the decode-and-forward rate almost coincides with the cut-set capacity upper bound, and it is close to the $8\times 4$ MIMO capacity when the relay is at $(d_x,d_y)=(0,\nicefrac{1}{10})$.
The highest decode-and-forward rate is attained when the relay is located approximately midway between the source and destination.
However, as the relay moves from the source and approaches the destination, the decode-and-forward rate begins to deteriorate.
In fact, when the relay enters into proximity of the destination, the decode-and-forward rate underperforms direct transmission.
This is because the decode-and-forward scheme requires the relay to fully decode the message from the source, and consequently the source-relay channel becomes a performance bottleneck.
In practice, the source would typically enlist the relay's help only if it offers a capacity gain over direct transmission;
hence the achievable rate may be taken as the maximum of the relaying rate and the direct rate.

The half-duplex decode-and-forward scheme follows similar trends, but exhibits a wider gap from the half-duplex cut-set capacity upper bound, and its maximum capacity gain over the direct channel is less pronounced.
In Fig.~\ref{fig:relay_dist01_BW_Mi4_R50}, it is observed that the system bandwidth is disproportionately allocated to Band~1 when the relay is in the vicinity of the destination, which corroborates with the system performance being limited by the source-relay link.
The orthogonal two-hop relaying scheme, on the other hand, does not perform as well as decode-and-forward.
It only offers moderate capacity gain over direct transmission, where the gain similarly is at its peak when the relay is about equidistant from the source and destination.

%%% ============================================================
\section{Compress-and-Forward Relaying}
\label{eq:CF_relaying}

From discussion in the previous section, it is clear that requiring the relay to decode the source's message can become a performance bottleneck when the source-relay channel is weak.
In this section, we consider the compress-and-forward \cite[Thm.~6]{cover79:cap_relay} strategy where the relay does not attempt to decode the message from the source.
Rather, the relay forwards a compressed version of its observation to the destination.
The relay's observation is compressed in the sense that a finite number of bits is used to represent the analog signal
(the scheme is sometimes also referred to as quantize-and-forward).
Unlike the cut-set bound and decode-and-forward formulations, however, the transmit signal design and bandwidth allocation under compress-and-forward do not appear to be convex problems.
In this section, we consider achievable compress-and-forward transmission schemes.
We focus on full-duplex transmission; under fixed bandwidth allocation, the compression-and-forward operation readily extends to half-duplex relaying.

\subsection{Compress-and-Forward Transmission}
\label{sec:CF_transmission}

We first describe the general compress-and-forward strategy;
specific compression schemes are considered in Sections~\ref{sec:CF_RD_compress}, \ref{sec:CF_WZ_compress}.
The optimal joint design of the transmit signals and compression rate appears to be intractable;
in the following we present suboptimal approaches to consider specific power allocation and compression schemes.
We assume the source and the relay use Gaussian signals.
Using the capacity-achieving strategy as in a multiple-access channel \cite{cover91:eoit},
suppose the destination performs successive interference cancellation to allow simultaneous transmission from the source and relay.
In particular, we consider the decode order in which the destination first decodes the relay's message, treating transmission from the source as noise.
Then the relay's codeword is subtracted from the observed signal, and the message from the source is decoded.
The source-destination transmission is thus interference-free from the relay's signals, and the source optimizes its own transmit signal covariance $Q_{11}$ according to
\begin{align}
\label{eq:CF_R_11}
R_{11} &= \max_{Q_{11}\;:\;\tr Q_{11} \leq P_1} \;\log\det\bigl( I_{N_1} + H_{11}Q_{11}H_{11}^H \bigr)
\end{align}
where the solution is given by the waterfilling procedure.
Let $Q_{11}^*$ denote the covariance matrix that maximizes (\ref{eq:CF_R_11}).
Next, the relay optimizes its transmit signal against the interference from the source's transmission
\begin{align}
\label{eq:CF_R_12}
R_{12} &= \max_{Q_{22}\;:\;\tr Q_{22} \leq P_2} \;\log\det\bigl( I_{N_1} + \tilde{H}_{12}Q_{22}\tilde{H}_{12}^H \bigr)
\end{align}
where $\tilde{H}_{12}$ is the effective channel from the relay to destination treating interference from the source as noise
\begin{align}
\tilde{H}_{12} &\triangleq (I_{N_1}+H_{11}Q_{11}^* H_{11}^H)^{\nicefrac{-1}{2}}H_{12}.
\end{align}
Similarly, the solution in (\ref{eq:CF_R_12}) is given by waterfilling against the effective channel $\tilde{H}_{12}$.

In the compress-and-forward approach, the relay sends $\tilde{y}_2 \in\mathds{C}^{N_2}$ to the destination, which is a compressed version of the relay's receive signal $y_2$, with compression rate $R_{12}$ as given in (\ref{eq:CF_R_12}).
The compression schemes considered in this paper can be modeled by
\begin{align}
\label{eq:yh2_Ay2_zt}
\tilde{y}_2 = A y_2 + \tilde{z}
\end{align}
where $A\in\mathds{C}^{N_2\times N_2}$ is a constant scaling matrix,
and $\tilde{z}\sim\mathcal{CN}(0,Z) \in\mathds{C}^{N_2}$ is independent additive Gaussian compression noise, with $Z\in\mathds{H}_+^{N_2}$.
Upon receiving $\tilde{y_2}$ at the destination, the relay network is equivalent to an $M_1 \times (N_1+N_2)$ MIMO channel,
except that $N_2$ of its receive antennas are scaled by $A$ and corrupted by compression noise $\tilde{z}$
\begin{align}
\label{eq:y1_yh2_H_A_D}
\begin{bmatrix}
y_1\\ \tilde{y}_2
\end{bmatrix} &=
\begin{bmatrix}
H_{11}\\ A H_{21}
\end{bmatrix} x_1 +
\begin{bmatrix}
z_1\\ A z_2 + \tilde{z}
\end{bmatrix}.
\end{align}
The corresponding channel capacity is given by
\begin{align}
R_{\CF} &= \log\det\bigl(I_N + \hat{H}_1Q_{11}^*\hat{H}_1^H \bigr)
\end{align}
where $\hat{H}_1$ is the effective source-to-destination-and-relay MIMO channel, incorporating the degradation introduced by the compression scheme as given in (\ref{eq:y1_yh2_H_A_D})
\begin{align}
\hat{H}_1 \triangleq
\begin{bmatrix}
H_{11}\\
(Z+AA^H)^{\nicefrac{-1}{2}} A H_{21}
\end{bmatrix} \in \mathds{C}^{N \times M_1}.
\end{align}
The different compression schemes considered in this paper differ in their respective achieved values of $A$ and $Z$.
They are described in more detail in the next sections.

\subsection{Rate--Distortion Compression}
\label{sec:CF_RD_compress}

Let $\hat{y}_2 \in\mathds{C}^{N_2}$ represent the compressed version of the signal $y_2$ under rate--distortion theory \cite{cover91:eoit}.
Let $\hat{z} = y_2-\hat{y}_2$ be the compression error, where $\hat{z}$ has zero mean.
The quality of the compression is characterized by the distortion covariance matrix: $D\in\mathds{H}_+^{N_2}\triangleq\E[\hat{z}\hat{z}^H]$.
In general, we wish $D$ to be small to obtain a high compress-and-forward rate.
In the following, we consider a simple approach to model this dependency by considering minimizing: $\tr D$,
which represents the sum of the quadratic distortion measures in the entries of $\hat{y}_2$.
The distortion-rate function prescribes the minimal distortion achievable under the compression rate $R_{12}$
\begin{align}
\label{eq:min_trD_Iy_R12}
\min_{D\;:\;I(y_2;\hat{y}_2) \leq R_{12}}\, \tr D.
\end{align}
After evaluating the mutual information expression in (\ref{eq:min_trD_Iy_R12}) under Gaussian signaling,
the minimization in the distortion-rate function can be written as
\begin{align}
\label{eq:RD_tr_D}
\text{minimize}\quad & \tr D\\
\label{eq:RD_over_D}
\text{over}\quad & D\in\mathds{H}_+^{N_2}\\
\text{subject to}\quad
& \log\det (S_2-D) \leq R_{12}.
\end{align}
Let $D^*$ denote the optimal distortion covariance matrix in (\ref{eq:RD_over_D});
it is computed by the reverse waterfilling \cite{cover91:eoit} procedure along the eigenmodes of $S_2$.
The $\hat{y}_2$ that achieves the minimum sum distortion in (\ref{eq:RD_tr_D}) has a joint distribution with $y_2$ described by
\begin{align}
\label{eq:y2_yh2_zh}
y_2 &= \hat{y}_2 + \hat{z},& \hat{y}_2 &\sim \mathcal{CN}(0,S_2-D^*),& \hat{z} \sim \mathcal{CN}(0,D^*).
\end{align}
The joint distribution in (\ref{eq:y2_yh2_zh}) can be equated with the form in (\ref{eq:yh2_Ay2_zt})
by setting the corresponding parameters of the compression scheme to be
\begin{align}
A_{\RD} &= (I_{N_2} - D^* S_2^{-1})^{\nicefrac{1}{2}},&
Z_{\RD} &= D^*
\end{align}
where the subscripts in $A_{\RD},Z_{\RD}$ are used to designate the compression scheme under consideration.

\subsection{Wyner--Ziv Compression}
\label{sec:CF_WZ_compress}

In the compress-and-forward strategy in \cite[Thm.~6]{cover79:cap_relay},
the transmission scheme also takes advantage of the correlation between the observed signals at the source and relay using Wyner--Ziv compression.
In particular, when the destination attempts to reconstruct $y_2$ form $\tilde{y}_2$, it also has access to its own receive signal $y_1$, which can be used to improve the performance of the compression process.
The Wyner--Ziv compression approach \cite{wyner78:rate_dist_si_gen} exploits the correlation between $y_1$ and $y_2$ as side information at the decoder
to achieve a lower compression noise level with the same compression rate $R_{12}$.

With the transmit signal of the source being as specified in (\ref{eq:CF_R_11}),
the covariance of the observed signals at the destination (after successive interference cancellation of the relay's signal) and relay, respectively, are given by
\begin{align}
S_{11} \triangleq \E[y_1y_1^H|x_2] &= I_{N_1} + H_{11}Q_{11}^*H_{11}^H \; \in \mathds{H}_+^{N_1}\\
S_{22} \triangleq \E[y_2y_2^H] &= I_{N_2} + H_{21}Q_{11}^*H_{21}^H \; \in \mathds{H}_+^{N_2}.
\end{align}
Moreover, the cross-covariance between $y_2$ and $y_1|x_2$ is
\begin{align}
S_{21} &\triangleq \E[y_2y_1^H|x_2] =  H_{21} Q_{11}^* H_{11}^H \; \in \mathds{C}^{N_2\times N_1}.
\end{align}
For Gaussian signals under quadratic distortion, the Wyner--Ziv scheme achieves the same rate--distortion tradeoff as if the side information were also present at the encoder \cite{wyner78:rate_dist_si_gen, zamir96:rate_loss_WZ}
(i.e., as if the relay had access to $y_1$ in the course of the compression process).
Therefore, the Wyner--Ziv compression noise is given by the distortion-rate function of compressing the signal $y_2|y_1$ using rate $R_{12}$
\begin{align}
\text{minimize}\quad & \tr \bar{D}\\
\label{eq:WZ_over_Db}
\text{over}\quad & \bar{D}\in\mathds{H}_+^{N_2}\\
\text{subject to}\quad
& \log\det (S_{2|1}-\bar{D}) \leq R_{12}
\end{align}
where $\bar{D}$ represents the distortion covariance matrix of the Wyner--Ziv compression error, and
$S_{2|1}$ is the conditional covariance of $y_2$ given $y_1$
\begin{align}
S_{2|1} \triangleq \E[y_2y_2^H|y_1]
=  S_{22} -  S_{21} S_{11}^{-1} S_{21}^H \; \in \mathds{H}_+^{N_2}.
\end{align}
The parameters for the Wyner--Ziv compression scheme are then identified to be
\begin{align}
A_{\WZ} &= (I_{N_2} - \bar{D}^* S_{2|1}^{-1})^{\nicefrac{1}{2}},& Z_{\WZ} = \bar{D}^*
\end{align}
where $\bar{D}^*$ is the solution to (\ref{eq:WZ_over_Db}) from reverse waterfilling against $S_{2|1}$.
Since the side information reduces the compression noise, Wyner--Ziv compression always achieves better performance than the rate--distortion compression scheme.
However, rate--distortion compression has lower implementation complexity,
since
the correlation between the receive signals of the source and relay is not exploited in the compression process.

\subsection{Numerical Results}
\label{sec:cf_num_res}

\begin{figure}
  \centering
  \includegraphics*[width=10.85cm]{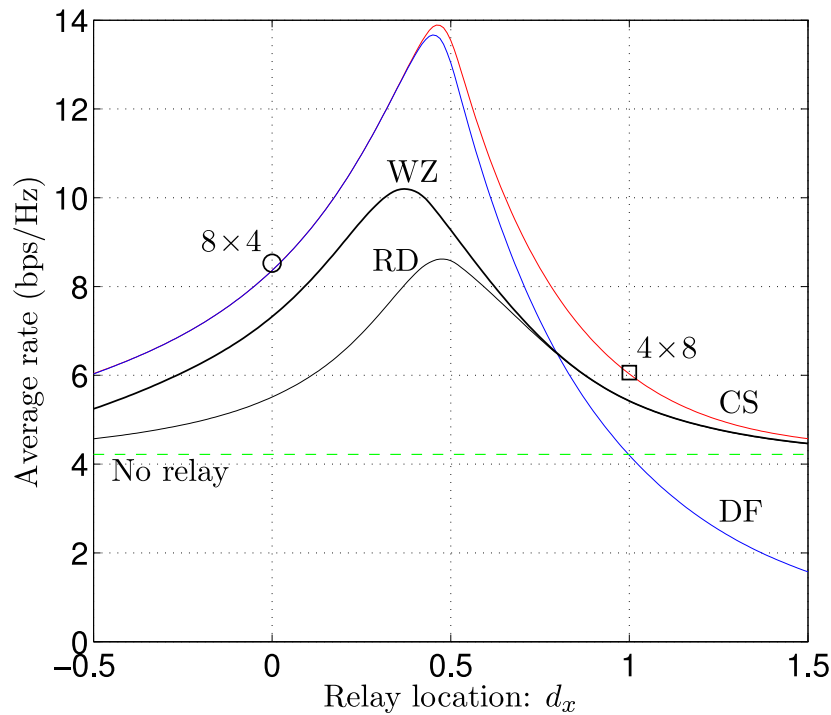}
  \caption{Compress-and-forward relaying rates under Wyner--Ziv (WZ) compression with side information and the rate--distortion (RD) compression scheme.}
  \label{fig:relay_dist01_CF_Mi4_R50_co}
\end{figure}

Fig.~\ref{fig:relay_dist01_CF_Mi4_R50_co} shows the compress-and-forward rates under the same channel parameters as those considered in Fig.~\ref{fig:relay_dist01_Mi4_R50_co}.
The Wyner--Ziv (WZ) compress-and-forward rate outperforms the compression-and-forward rate under rate--distortion (RD),
which demonstrates the capacity gain from exploiting side information.
However, the Wyner--Ziv advantage ceases when the relay is close to the destination:
in that regime, the efficiency of the compression scheme has limited impact, as the relay has a strong channel to the destination.
Overall, the compress-and-forward rates do not perform as well as the decode-and-forward rates, except when the relay is far from the source and near the destination.
Moreover, unlike its decode-and-forward counterpart at $(d_x,d_y)=(0,\nicefrac{1}{10})$,
the compress-and-forward rates markedly fall short of the $4\times 8$ MIMO capacity when the relay is at $(d_x,d_y)=(1,\nicefrac{1}{10})$.
Nevertheless, as the relay is under no stipulation to perform any decoding, the compress-and-forward rate is at least as large as that under direct transmission,
regardless of network geometry.

%%% ============================================================
\section{Conclusions}
\label{sec:conclu}

We considered the optimization of transmit signals and bandwidth allocation for MIMO relay channels.
We assumed that all terminals have channel state information,
and we evaluated the cut-set capacity upper bounds and the decode-and-forward rates by formulating them as convex optimization problems.
The solutions to the optimization problems can be efficiently computed by numerical convex optimization methods.
In the case of half-duplex relaying, where the relay cannot simultaneously transmit and receive in the same frequency band,
the bandwidth allocation and the transmit signals are jointly optimized.
We also presented achievable relaying rates based on the compress-and-forward strategy, where the relay does not decode the message from the source,
but forwards a compressed version of its observation to the destination using the rate--distortion and Wyner--Ziv compression schemes.

When the relay is close to the source, it is observed that the decode-and-forward coding strategy is almost optimal:
its achieved rate is near the cut-set capacity upper bound, especially in full-duplex relaying.
Moreover, under the half-duplex constraint, decode-and-forward significantly outperforms orthogonal two-hop relaying.
For all relaying schemes, the maximum capacity gain over direct transmission is attained when the relay is approximately halfway between the source and destination.
On the other hand, when the relay is close to the destination, decode-and-forward underperforms direct transmission as the source-relay link becomes a bottleneck.
In this regime good performance is achieved by the compress-and-forward schemes, which always achieve a rate that is equal to or better than the direct transmission rate.

%%% ============================================================
\bibliographystyle{IEEEtran}
\bibliography{IEEEabrv,wrlscomm}

\end{document}